# Minimum amount of energy required for the transference of electrons between atoms

*Pillon, Matheus Barbiero*
*Institute of Bom Jesus Ielusc*
*Joinville, SC, Brazil*

## Abstract

*The purpose of this article is to show a different way to calculate the electron transfer between atoms. It can be used as a base for calculating the minimal electricity, showed here by Joules, necessary for the transference of electrons. There is already a Table, in which you find the result of all of the electric transfer necessary in most of the materials. In this article, we tried to generalize the electricity necessary, but if you want the result in a specific field or material, you can search the references and collaborates workers.*

## 1. Introduction

Quantum physics, also known as quantum mechanics, is one of the newest areas of physic. Created in 1900 by Max Planck, quantum mechanics is known as one of the hardest subjects to learn and work with. The field deal with the behavior of matter and energy in the scale of atoms and subatomic particles/waves [2]. The quantum physic, seen as a difficult subject, is used every day by everyone. It can be regarded in actions as charging your cellphone, turning on your microwaves, or even clicking on the red bottom to watch your TV [3] [4].

One of the subareas of quantum mechanics is electrodynamics, which is the central speech of this article. Already worked by famous scientists as *Richard Feynman, Sin-Itiro Tomonaga and Julian Schwinger*, the subarea deal with the reactions of the electron or radiative correlation on precisely quantum phenomena [5]. The main idea of this article is the minimal purposeful transfer of electrons between atoms. It proposes the minimal amount of energy necessary to send the electron from one atom to other.

This article is divided into four parts: Introduction, Collaborates workers, Proposal, and Conclusion. Each part has details about the work's ideas and ways to understand and prove them.

## 2. Collaborates workers

### 2.1 General ideas

One of the main ideas of this article is based on the works of Bohr, in which he proposes one formula (Equation 1) that calculates the amount of energy necessary for changing electrons between atoms axes. I used it as a principle for creating the formula that cares about the energy necessary for the achievement of an electron on a different atom [1].

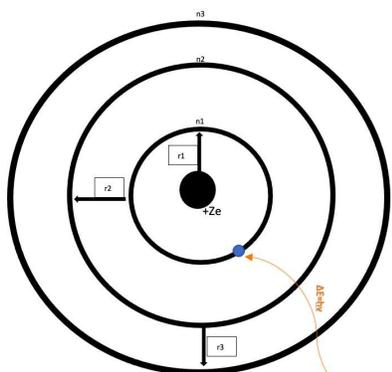

*Figure 1: The model made by Niels Bohr. In the image, it shows the axes from the atom, represented by the letter **n**. The letter **Z** is used as a representation of the atomic number. The positive signal (+) and the letter **e** are showing that the atom is ionized, which means that it has fewer electrons in its axes than normal.*
*Based on Niels Bohr* [1]

In his work, Niels Bohr does speak about the changing of electrons in the same atom in different axes, shown in figure 1, but he has never spoken about the transfer of electrons between atoms, which is the main idea of this article (Figure 1) [1].

One example of this technology nowadays is showed by professors at the Duke University and the University of Cyprus, named *David N. Beratan, Spirios S. Skourtis, Ilya A. Balabin, Alexander Balaeff, Shahar Keinan, Ravindra Venkatrammi, Dequan Xiao*, in which they speak about the consequences of the electron transfer reaction in proteins, chemistry and biology areas. They show and calculate, with graphics and images, how is the reaction and how the protein is modified with electron transfer [6].

We could do these works with the formula (Equation 2) showed in this article, which would help with the energy wasted and also the conduct of this observation in laboratory.

Another group of physicians that speaks about electron transfer is *Ziegler and Biersack* (1985). In their article, they speak about how calculating the stopping and the final range distribution of ions in matter. They use graphics and tables, in which they show the evolution of the ionized atoms stopping. It is also calculated in the article, the hydrogen mass, and energy lost on the stopping [7]. They have used their formula, which contains more difficult terms than mine, but, on the other side, their formula is more appropriate for their work, which requests more detailed and approximate results [7]. It can be also cited here, the article of *Cazaux*, which speaks about the ionizing radiation in electrodynamics, with more precise, the changing effects of transmission electron microscopy, with formula, figures, and graphics [8]. In his formula, there is the presence of the integral, because of the electrical transmission changing. He deals with the temperature changing and more than one angle field [8].

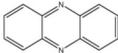

*Table 1: It is shown here the section of exogenous redox mediators used for microbial fuel cells. In other words, it is a section of the microbial fuel cells, which are chemicals with an electrochemical activity that are growing or creating themselves outside of an organism or part of it. This table is showed in Uwe Shröder's article.*

*Font: Uwe Schröder* [9]

On other hand, his formula has a lot of terms and deals with the changing, but not with a static moment, which is carried by my formula [8].

## 2.2 Specifics areas

There are also a lot of examples in specifics areas that wouldn't handle in this article, even though I will cite some of them in different fields: *Uwe Schröder* has shown in his article in April of 2007 solutions of the problem to the electron transfer between microscopy beings and has also proposed the energy efficiency of the anodic electron transfer mechanism. For his work, he needed to measure the electron's energy between the two microbial full cells, which he could have used my formula [9].

Besides, his job is related to mine by the electron transfer and the electrodynamics, even so, he uses it in a subarea, where is more important the details that he has achieved with the *Table 1* and the *Table 2* [9].

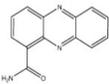

*Table 2: It is the comparative of the Table 2. While the Table 2 has the information about the section of exogenous redox mediators used for microbial fuel cells, this table shows the selection of extracellular bacterial (endogenous) redox mediators. It was used by Uwe Shröder for his research, but what is the difference between the two tables? While the Table 2 speaks about exogenous microbial fuel cells, which means antigens entering from outside of your body to inside of your body, and this table tells us about Endogenous, which means antigens that are already inside of your organism and are causing problems.*

Font: *Uwe Schröder* [9]

The article of *Yan Jiao, Yao Zheng, Mietek Jaroniecb, and Shi Zhang Qiao* promise the Design of electrocatalysts oxygen- and hydrogen-involving energy conversion reactions. Their article's main idea is to study one way to reduce the climate change due to the energy conversion reaction [10].

Their works are related to mine because of the energy conversion reaction, which is carried by electron transfer. The article works with specific areas, because of the hydrogen and oxygen-involving energy and because of being treated in the chemicals field [10].

Another chemistry article, in which it is spoken all of the electron transfer reaction in this area was written by *Rudolph A. Marcus*. He tells us about all of the discoveries since the 1940s and explains how to solve problems with some specific applications and exceptions showed in these articles [11].

## 3. Proposal

### 3.1 Conditions

Before we begin the explication, we need to show the condition for the function of the formula. In other words, I'll show the necessary hypotheses. Opening, we need to consider that the two atoms are in a straight line and the electron needs to be static, in other words, with no speed. This is necessary for the function of the formula, with this, the electron will go in a straight line between the atoms. This lets us see the next condition, we need to consider that the electron follows the same steps and the same compartments inside and outside the atoms. Without that, the electron couldn't be transferred to the next atom.

$$Ep = \frac{E1}{n^2}$$

*Equation 1: This is the formula of Bohr, in which he showed how to measure the potential energy of each electron on a specific axis. The letter "n" is the number of the axis and the letter "E" is the energy, in this case, E1 is $-2.18 \times 10^{-18}$.*

*Font: Niels Bohr* [1]

Also, we need to suppose that we can measure the distance between atoms. If this is not possible, the whole idea of this article is false.

Finally, we need to set the limits of the atom. As Niels Bohr has shown us, the potential energy of an electron on axis one is $2.18 \times 10^{-18}$ J, and, with his formula Equation 1, we can measure all of the potential energy of the electron on all of the axes from the atom [1]. With this in mind, we can say that the minimum energy necessary is more than $2.18 \times 10^{-18}$, since the electron needs to be a positron, anti-electron, to be guided to his final destiny, the next atom. We know that this is the energy necessary for the electron transform itself, because of Bohr's formula Equation 1, which shows us the potential energy of the electron, in this case, $2.18 \times 10^{-18}$, considering that the electron is on the first axis. It is also considered here that it needs to still use the kinetic energy for ruining in his straight line. If it is not possible, he will be stopped in the vacuum [1]. Even though there is minimal energy necessary, there isn't any limit, because this formula aims to transfer whatever electron at whatever distance, the only thing necessary is the measurement of this distance.

## 3.2 The Idea

The principle of this article, as the name suggests, is for searching for the minimal purposeful transfer of electrons between atoms. It suggests that, with the formula, we could save the electricity or use it differently way without put in more or less than necessary for the escaping of the electron in direction of another atom. We can apply it in laboratory on researches, in which we need a negative energy atom and, due to the lack of money or intending to save energy, we use the formula for finding what is the minimum necessary, showed in Equation 2.

$$Eat = \frac{d}{rf|(n^2 - (n+1)^2)|} \times |Ef| \div (n^2 - (n+1)^2)$$

*Equation 2: Energy's formula of minimal purposeful transfer of electrons between atoms*

The letter "d" represents the distance between the two atoms, while the letter "n" represents the number of the axis the electron is. The "fs" after "E" and "r" mean fundamental, which is necessary for the functioning of the formula in more than one atom, you just need to find the fundamental energy and radius.

The module on the "Ef" is necessary because the electron needs to have positive energy to transform itself on a positron, anti-electron, and get out of the atom. The module is also necessary on the parentheses that are multiplied by "rf", since the distance is never negative if there is progression. This formula can also be reduced with just three points, shown in Equation 3.

$$Eat = \frac{D}{ad} \times aE$$

*Equation 3: Energy's formula of minimal purposeful transfer of electrons between atoms in parts*

The term "Eat" is named that way, because of its function. It means "Electric atomic transfer", while the terms "ad" and "aE" mean axis distances and axis energy or axis potential energy. The isolated letter "D" tells us about the distance between the atoms. The calculus made in this article uses the hydrogen atoms, which the fundamental energy is $2.18 \times 10^{-18}$ and the fundamental radius is $5.3 \times 10^{-11}$, which are carried by the first axis. For finding this information, you can search the table from Niels Bohr on the reference [1].

### 3.3 The Examples

For the understanding of the formula, we can imagine one ordinary action where most of us normally ask in a day by day. You need to go to the supermarket, where you will do your monthly shopping, but you live on a dangerous street, because of that, you should go with a car, you don't have one, consequently, you take your brother's car to go. When you join into the car, which is garbage near a lightbulb, you find out that the car has less than one eighth of your fuel. You know the distance between your house and the supermarket, since you have opened your *GPS* and set your address and the location where you want to go, in this case, supermarket.

The car that you are using is your brother's automotive, by consequence of that, you do not know how much it loses per kilometer. How can you know if you will or not go to the supermarket without needing to call an engineer because of the lack of fuel? In this situation, you will probably call your brother and ask for the information or simply don't go to the supermarket, but as this is an example, we cannot do such things. You drive from the lightbulb, where your brother's car is garbage, to the door which you use for joining the street. When you achieve the door, you discover that you have lost 100 ml of gasoline.

Subsequently, you drive until you achieve the first car garaged in your street and you see that you lost, in total, 150 ml of gasoline. You subtract the two losses of gasoline and find out that the fuel's lost between the distance of your garage and the first car garaged in your street is 50 ml.

Knowing, due to the *GPS*, that there are more than 50 meters to your destiny, and you drove just 5 meters, you know that the total gasoline needed to the supermarket and the first car garaged in your street is 500 ml. Let's try to use the formula for calculating the example:

$$\frac{50}{|5|} \times |100 - 150|$$
$$\frac{50}{5} \times 50$$
$$10 \times 50$$
$$500$$

You have probably noted that I have used the module on this calculus, which are not used on the formula (Equation 2). This is necessary in this case, because the result would be negative, without it, the result wouldn't be correct.

Even though, on an atomic scale, with more distance of the proton, more have the electron positive energy, which means, bigger number of axis, less negative energy. If you want a more scientific explication, I shall show with images, for a better understanding.

We see in the Figure 2 that the atom A is a hydrogen atom, due to its isolated proton, and it has an

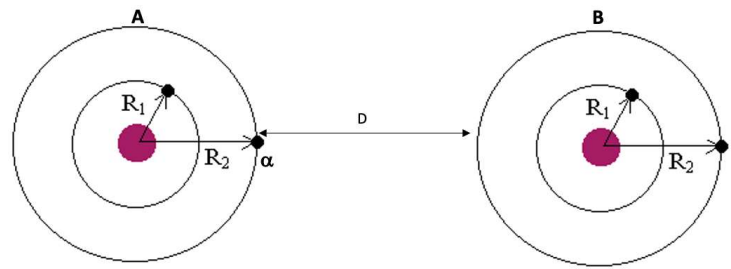

*Figure 2: The first atom is represented by the letter "A", while the atom, which the electron will be transferred, is represented by the letter "B". The letter that represents the electron that will be transferred is "α"*

*Font: The Author.*

electron. Let's imagine that we want to throw the electron α into the atom B. For this to be happen, we need to first know which axis the electron is, in this case, it is on axis two.

The second step is to discover the next axis. For example: axis 2+1= axis 3. The third step is finishing the algebra from the formula with the information you already have.

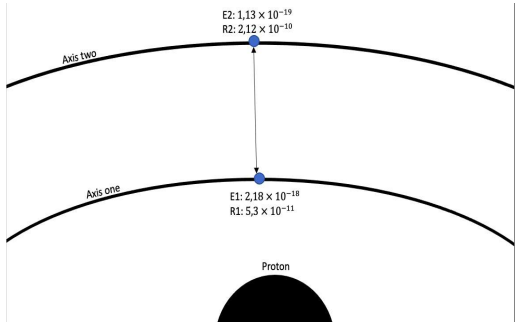

Continuing, you just have to find what is the distance between the axis from the electron α and the next axis.

Then, you'll finish the algebra of the formula. After finishing the part "ad", the distance between axis, and "aE", the energy between axis, you need to divide the part "ad" by D, the distance between atoms.

Finally, multiplying the result by the party "aE", getting the final result.

In Figure 3, there is a closer picture of the distance of the axis one and two, which can be used for the understanding of the formula.

*Figure 3: It is shown here the potential energy from the axis one and axis two and the radius from the axis one and two to the proton, cited for the understanding of the idea of subtracting between the two axes and why used you use it. The two blue points are the electrons, while the big black point is the proton. With this in mind, you can already guess that this is a representation of a hydrogen atom, but the formula can also be used with different atoms.*

*Font: The Author*

You are probably asking why I would have to calculate the distance between the axis if you want to know the energy between the atoms. This is necessary because of the idea of the formula, I have used a mechanism that takes the distance between the axis and divide all over the distance between the atoms, with that you just need to transcribe the energy. The formula seems to be kind of difficult, but the idea is quite simple. The only thing that makes it look difficult is the Bohr's formula that is needed in these cases [1].

### 3.4 Practicing with Joules

For the best knowlegdge of the formula, I will use it with imaginary numbers for the example. We can suppose that the distance between the atom with electrons and the atom we want to transfer is 1nm and the axis the electron stayed is the number three. With the formula, we first discovered the subtraction of the axis, in which the electron stayed, and the next axis.

$$Eat = \frac{D}{5.3 \times 10^{-11}|(n^2 - (n+1)^2)|} \times 2.18 \times 10^{-18} \div (n^2 - (n+1)^2)$$

$$Eat = \frac{D}{5.3 \times 10^{-11}|(3^2 - (3+1)^2)|} \times \frac{2.18 \times 10^{-18}}{9} - \frac{2.18 \times 10^{-18}}{16}$$

$$Eat = \frac{D}{5.3 \times 10^{-11}|(9-16)|} \times \frac{34.88 \times 10^{-18}}{144} - \frac{19.62 \times 10^{-18}}{144}$$

$$Eat = \frac{D}{5.3 \times 10^{-11} \times 7} \times 15.26 \times 10^{-18}$$

$$Eat = \frac{D}{37.1 \times 10^{-11}} \times 15.26 \times 10^{-18}$$

$$Eat = \frac{1}{37.1 \times 10^{-11}} \times \frac{15.26 \times 10^{-18}}{1}$$

$$Eat = \frac{15.26 \times 10^{-18}}{37.1 \times 10^{-11}}$$
$$Eat = \frac{15.26}{37.1} \times 10^{-7}$$

As we see, the Joules necessary from the axis three to four is $15.26 \times 10^{-18}$. Continuing our calculus, we shall look for the multiplication between $15.26 \times 10^{-18}$ and $1/37.1 \times 10^{-11}$, getting the result of $\frac{15.26}{37.1} \times 10^{-7}$ J necessary for this distance.

### 3.5 Exceptions

There are always exceptions, in which there are problems that in some circumstances the formula won't work, however, there are also problems where, with the adaption of the formula, the same one can be used.

In their article, *William A. Tisdale, Kenrick J. Williams, Brooke A. Timp, David J. Norris, Eray S.* and *Aydil, X.-Y. Zhu* have shown the efficiency of nanocrystals in hot atoms. They have shown the solution of high-energy photons transfer, but they have not mentioned what could be done for solving the electron transfer [12].

In these conditions, it is already known that one of the solutions is the addition of a variant "T" on the formula, but, more than that, the formula needs to be restored. Knowing that, with the solar cells, the photons are reduced by ~31%, cited by Tislade, the electron transfer could also change and knowing that it cools fast (~1 ps), also cited by Tislade, it would be necessary an equation with variation, including an integral. In whatever way, this is not the goal of this article, it will be carried in future works [12].

In a different scenario, for example, the uses of artificial atoms, how could the electrons behave? The answer is shown by [13] [14].

The behave of the electron is not that different, even so, one detail is quite important: Some barriers are allowed for being cross on artificial's atoms. This means that we need to be careful about which field and which interferences: pressure, temperature, etc., the two atoms that we chose that will be the ones are [14] [13].

To find the solution for this problem, we will need to search for the conditions of the place and then think about the changes that will need to be done. What could be done in futures articles, knowing that this is not the goal of this article.

## 4. Conclusion

In conclusion, we have seen in this article references and formulas for the minimal purposeful transfer of electrons between atoms. It can be used by radiation conducted in laboratory and it can also be used in futures research. It is pendent here what the electron will do if there are interferences of high or low temperature or high or low pressure between atoms, where the formula does not work or even if there are possibilities that with interference, the electron couldn't, with the minimal amount of energy calculated by the formula, achieve its destiny. The problem with the temperature could be solved with the variant "T", but it is not much necessary if there aren't many changes in the field, for example, a well-controlled laboratory.

It is also pendent in this article, the result with details and specific information in a specific material or field. The formula can be useless if you are working in an area where you have to show non-approximate results or a lot of details about electron transfer. Finally, all of the questions that I have cited in the *Exception* section, which I have already said the possible solution of the problems.